\documentstyle[prl,aps,graphics,epsf,twocolumn]{revtex}

\begin{document}

\title{Precision frequency measurement of visible intercombination lines of strontium}

\author{G. Ferrari, P. Cancio$^{1}$, R. Drullinger, G. Giusfredi$^{1}$, N. Poli,\\ M. Prevedelli$^{2}$, C. Toninelli, and G.M. Tino}
\address{Dipartimento di Fisica and LENS, Universit\`a di
Firenze, INFM-UdR Firenze, Polo Scientifico, 50019 Sesto Fiorentino, Italy\\
1) Istituto Nazionale di Ottica Applicata, Largo E. Fermi 2, 50125 Firenze, Italy\\
2) also Dipartimento di Chimica Fisica, Universit\`a di Bologna, Via del
Risorgimento 4, 40136 Bologna, Italy}

\date{\today}
\maketitle

\begin{abstract}
We report the direct frequency measurement of the visible
5s$^2$\,$^1$S$_0$-5s\,5p$^3$P$_1$ intercombination line of strontium that is
considered a possible candidate for a future optical frequency standard. The
frequency of a cavity-stabilized laser is locked to the saturated fluorescence
in a thermal Sr atomic beam and is measured with an optical-frequency
comb-generator referenced to the SI second through a GPS signal. The $^{88}$Sr
transition is measured to be at 434\,829\,121\,311\,(10)\,kHz. We measure also
the $^{88}$Sr-$^{86}$Sr isotope shift to be 163\,817.4\,(0.2)\,kHz.
\end{abstract}

\vspace{5mm}


\vspace{5mm}

\begin{figure}[h]

\epsfxsize=8cm \centerline{\epsfbox{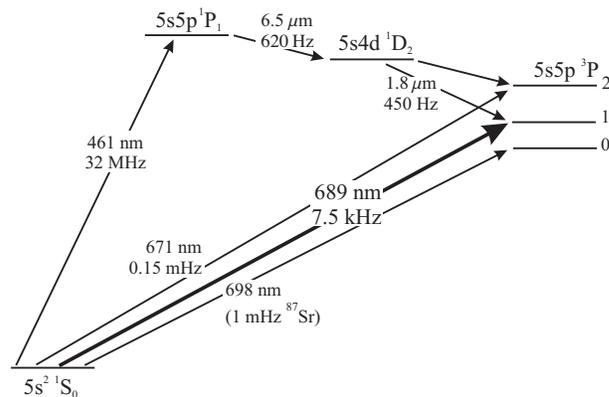}}
\caption{\label{LevelScheme} Relevant energy levels (not to scale)
and transition linewidth for high resolution spectroscopy and
atomic manipulation of bosonic strontium.}
\end{figure}

The recent development of optical-frequency comb generation has made possible,
for the first time, relatively easy optical frequency measurements
\cite{Udem99,Diddams2000}. This, in turn, opened the way to atomic clocks based
on optical frequency transitions. Because of their higher frequency, these
transitions have potential for greatly improved accuracy and stability relative
to conventional atomic clocks based on microwave frequency transitions
\cite{justifocation optical clocks}. Different transitions are now considered
as optical frequency standards, involving single ions and neutral atoms
\cite{HgCaClock}. While single ions offer an excellent control on systematic
effects, clouds of laser cooled atoms have the potential for extremely high
precision. Amongst the neutral atoms, Sr has long been considered one of the
most interesting candidates \cite{Hall89}. Several features, some of which are
specific to this atom, allow different possibilities for the realization of a
high precision optical clock. The intercombination 5$^1$S-5$^3$P lines from the
ground state are in the visible and easily accessible with semiconductor lasers
(Fig.\ref{LevelScheme}). Depending on the specific fine-structure component and
on the isotope - Sr has four natural isotopes, three bosonic, $^{88}$Sr (82\%),
$^{86}$Sr (10\%), $^{84}$Sr (0.5\%) with nuclear spin I=0 and one fermionic,
$^{87}$Sr (7\%) with I=9/2 - a wide choice of transitions with different
natural linewidths is possible. These span from the 7.5\,kHz linewidth of the
5$^1$S$_0$-5$^3$P$_1$ line, which is the subject of the present paper, down to
the highly forbidden 5$^1$S$_0$-5$^3$P$_{0,2}$ transitions. In $^{87}$Sr, the
presence of hyperfine mixing makes the 0-0 transition weakly allowed with an
expected natural width of about 1\,mHz.

From the point of view of laser cooling and manipulation, Sr has
several interesting features which are also important for the
final operation of a precise frequency standard: two-stage cooling
using the intercombination transition allows extremely low
temperatures and magneto-optical trapping; atoms can be trapped in
optical lattices with negligible shift of the optical clock
transition \cite{Katori00}. After initial laser spectroscopy
experiments based on wavelength metrology and discharges as atomic
sources \cite{Tino1994}, recently Sr has been the subject of
several experiments aiming to all-optical cooling down to quantum
degeneracy for bosonic and fermionic isotopes
\cite{Katori99,Xu03,SrTf03}, continuous atom laser
\cite{Katori00}, and detection of ultra-narrow transitions
\cite{Courtillot03,Katori03}. This atom is also considered
interesting for the understanding of its spectrum
\cite{Derevianko01} and for the investigation of cold collisions
\cite{Dinneen98,Julienne}.

In this paper, we report the first precision frequency measurements on the
intercombination 5$^1$S$_0$-5$^3$P$_1$ transition. Using a femtosecond laser
comb, we determine the absolute frequency of the transition for $^{88}$Sr and
$^{86}$Sr and a very accurate value for the isotope shift. The improvement by
several orders of magnitude with respect to previous data and the use of a
relatively simple and compact apparatus demonstrate the potentialities of this
system.

\begin{figure}[h]
\epsfxsize=8cm \centerline{\epsfbox{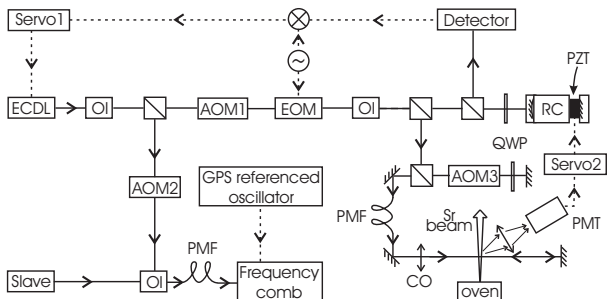}}
\caption{\label{SetupScheme} Experimental setup used for the
frequency measurement on the Sr intercombination line. Optical
isolators (OI) and acusto-optic modulators (AOM) eliminate
feedback among the master laser (ECDL), the slave laser, the
electro-optic modulator (EOM) and the reference cavity (RC). Solid
lines represent the optical path, dashed lines represent
electrical connections. QWP: quarter wave-plate. PMT:
photomultiplier tube. CO: collimation optics. PMF: polarization
maintaining fiber.}
\end{figure}

The experimental setup we use is composed of a laser-diode frequency-locked to
an optical cavity whose modes are locked to keep the laser on resonance with
the atomic line. The optical frequency is measured with a self-referenced
optical-comb stabilized against a Global Positioning System (GPS) controlled
quartz. A scheme of the experimental setup is given in Fig. \ref{SetupScheme}.
The extended cavity laser-diode (ECDL) is a Hitachi HL6738MG mounted in the
Littrow configuration which delivers typically 15\,mW. Optical feedback to the
ECDL is prevented by a 40\,dB optical isolator and a single pass acusto-optic
modulator in cascade. The laser linewidth is reduced by locking the laser to an
optical reference cavity (RC) with the classic Pound-Drever-Hall scheme
\cite{Pound-Drever-Hall}; the phase modulation is produced by a resonant
electro-optic modulator (EOM) driven at 21\,MHz. To avoid residual standing
wave in the EOM, which induces spurious AM on the locking signal, a 25 dB
optical isolator is placed between the EOM and the cavity. The reference cavity
has a free spectral range (FSR) of 1.5\,GHz and a finesse of 10000. On one side
of the quartz spacer we glued a concave mirror (R=50\,cm) while on the other
side a piezoelectric transducer (PZT) is glued between the spacer and a flat
mirror in order to steer the modes of the cavity by more than one FSR.

The lock of the laser onto the cavity includes a low frequency loop acting on
the PZT  of the ECDL (1\,kHz bandwidth), and a high frequency loop acting on
the laser-diode current supply (1\,MHz bandwidth). Under lock condition more
than 55\,\% of the incident light is transmitted through the cavity. From the
noise spectra of the locking signal and by comparison with another cavity we
can infer a laser linewidth less than 2\,kHz, and more than 90\,\% of the
optical power in the carrier \cite{ClaironStabilization}. We do not passively
stabilize the RC in a vacuum chamber \cite{UltraStableCavity} since the
acoustic and sub acoustic noise is removed by the servo to the atomic signal
which acts on the PZT of the RC with a 200\,Hz bandwidth.

The strontium atomic beam is obtained from the metal heated to 830\,K in an
oven and using a bundle of stainless steel capillaries to collimate it
\cite{Capillary}. The residual atomic beam divergency is 25\,mrad and the
typical atomic density in the detection region is $10^{8}$\,cm$^{-3}$.

The Doppler-free atomic line is resolved by saturation
spectroscopy using two counterpropagating laser beams
perpendicular to the atomic beam. The fluorescence light from the
laser excited atoms is collected on a photomultiplier tube with an
efficiency of 0.4\,\% including quantum efficiency and solid
angle. Orthogonality between atomic and laser beams is optimized
by centering the Lamb dip with respect to the Doppler profile.

The laser beam is filtered using a single mode fiber and
collimated at a $1/e^2$ diameter of 14\,mm (wavefront distortion
less than $\lambda$/6); the beam is retro-reflected using a mirror
at a distance of 65\,mm from the interaction region and coupled
back into the fiber. We estimate the indetermination on the angle
of the retroreflected beam to be less than 10\,$\mu$rad maximizing
the transmitted power through the fiber. The peak beam intensity
of 60\,$\mu$Wcm$^{-2}$ (to be compared to the saturation intensity
of 3\,$\mu$Wcm$^{-2}$) was chosen to obtain sufficient signal to
noise for the RC lock onto the atomic resonance. A uniform
magnetic field of 10\,G defines the quantization axis in the
interrogation region such that the light is $\pi$ polarized.

\begin{figure}
\epsfxsize=8cm \epsfbox{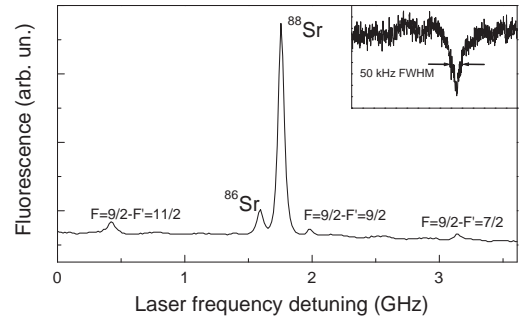} 
\caption{\label{BroadSpectrum} Fluorescence spectrum of the
strontium $^1$S$_0$-$^3$P$_1$ line at 689\,nm. The lines of the
two bosonic isotopes $^{86}$Sr and $^{88}$Sr, together with the
hyperfine structure of the fermionic $^{87}$Sr, can be resolved.
The linewidth corresponds to the residual $1^{\rm st}$ order
Doppler broadening in the thermal beam. Inset: sub-Doppler
resonance of $^{88}$Sr recorded by saturation spectroscopy using
two counter-propagating laser beams. The amplitude of the dip is
10\,$\%$ of the Doppler signal.}
\end{figure}

The acusto-optic modulators between the ECDL and the EOM (AOM1)
and between the ECDL and the slave laser (AOM2) are driven from
the same oscillator and both deliver the -1 order such that the
frequency instability and indetermination of their driving RF does
not affect the optical frequency measurement. The double pass AOM
next to the atomic detection (AOM3) is frequency modulated at
10\,kHz to derive the locking signal of the cavity onto the atomic
line.

Fig. \ref{BroadSpectrum} shows the Doppler broadened resonances of
$^{88}$Sr, $^{86}$Sr and the hyperfine structure of $^{87}$Sr. The
residual atomic beam divergency produces a residual Doppler
broadening of 60\,MHz FWHM. In the inset, the sub-Doppler signal
for $^{88}$Sr is shown. Two independent measurements
\cite{LaserLinewidth} of the sub-Doppler resonance show a FWHM of
about 50\,kHz, which is in agreement with the expected value
considering the saturation and transit time broadening, and the
recoil splitting.

\begin{figure}
\epsfxsize=8cm\centerline{\epsfbox{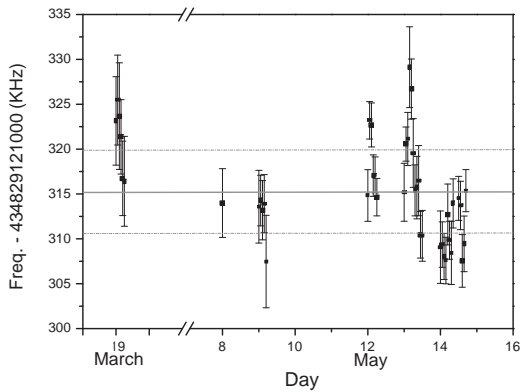}}
\caption{\label{Ripetibilita} Measurements used to determine the
transition frequency. The error bars correspond to the standard
deviation for each data set.}
\end{figure}

We measure the optical frequency through a commercial optical
frequency comb generator based on a Kerr-lens mode-locked Ti:Sa
laser with a repetition rate of 1\,GHz (MenloSystems GmbH, Model
FC8003). The repetition rate and carrier offset envelope frequency
are locked to a GPS stabilized quartz oscillator, as well as
counters and RF generators for AOM's. Figure \ref{Ripetibilita}
shows the result of the measurement of the $^{88}$Sr transition
frequency taken over a period of several days. Each data point
corresponds to the averaging of the values resulting from
consecutive measurements taken with a 1\,s integration time over
100-200\,s. The error bars correspond to the standard deviation
for each data set. The Allan deviation of each set shows a flicker
floor varying between 1 and 2\,kHz in the region from 1 to 100
seconds.

We evaluated 1$^{\rm st}$ and 2$^{\rm nd}$ order Doppler and Zeeman effects, AC
Stark shift, collisional shifts, and mechanical effects of light (table
\ref{TabBiasUncertainty}). The 1$^{\rm st}$ order Doppler shift resulting from
imperfect alignment in the standing wave was randomized by realigning the
retroreflected beam after each measurement; the resulting contribution in the
final uncertainty is included as 2\,kHz. The offsets and lineshape asymmetries
introduced by the recoil, atom deflection induced by the light field, and
2$^{\rm nd}$ order Doppler were calculated by numerically integrating the 1D
optical Bloch equations along the atomic trajectories considering the
experimental conditions \cite{Cancio03}. The resonance linewidth obtained from
this simulation is in good agreement with the experimental value proving that
we do not have unexplained line broadening mechanisms. Since we observe a
closed transition we estimate the offset introduced by unbalanced
counter-propagating beams and curved wavefront \cite{Borde76}, and wavefront
distortion less than 2\,kHz . There is no first order Zeeman shift because we
observe the J\,=\,0 to J\,=\,1, $\Delta$m\,=\,0 transition. The second-order
Zeeman shift in our magnetic field is of the order of few Hz. The collisional
shift coefficient for this transition has not been measured but the self
broadening coefficient is known to be about 50\,MHz/torr \cite{CollBroadening}.
Assuming as an upper limit for the collisional shift the self broadening
coefficient and considering the background pressure of the order of
10$^{-6}$\,torr, we expect a pressure induced shift of less than 50\,Hz.
Unbalanced sidebands on the interrogating laser  were measured to be more than
40\,dB below the carrier leading to a shift smaller than 1\% of the atomic
linewidth. In our experimental conditions, we did not experimentally observe
any dependence of the measured optical frequency on the modulation depth and
laser intensity, which is in agreement with numerical simulations. The
resulting value for  the $^{88}$Sr transition frequency, including the
corrections discussed previously, is 434\,829\,121\,311 (10) kHz, corresponding
to a $1\,\sigma$ relative uncertainty of $2.3\times 10^{-11}$.

With a minor change in the apparatus, we locked simultaneously the frequency of
two laser beams to the sub-Doppler signals of $^{86}$Sr and $^{88}$Sr. This
system allowed us to measure the isotope shift by counting the beatnote between
the two interrogating beams. For this purpose, the reference cavity is locked
to $^{88}$Sr resonance as described previously and the light for $^{86}$Sr is
derived from the same laser beam and brought to resonance through AOM's. The
two beams are overlapped in a single mode optical fiber and sent to the
interrogation region. By frequency modulating the beams at different rates and
using phase sensitive detection we get the lock signal for both the isotopes
from the same photomultiplier. The lock on $^{86}$Sr acts on the
voltage-controlled oscillator that drives one of its AOMs. The $^{86}$Sr lock
bandwidth of 1\,Hz, limited by lower signal to noise, is enough since the short
term stability is insured by the lock to the reference cavity and $^{88}$Sr. In
this isotope shift measurement most of the noise sources are basically common
mode and rejected; the Allan variance shows a white noise spectrum of 1\,kHz at
2\,s and does not show any flicker noise for times longer than 500\,s resulting
in precision better than 100\,Hz. At this level of precision we observe the
servo loop offset compensation limiting the reproducibility to 200\,Hz. The
measured $^{88}$Sr-$^{86}$Sr isotope shift for the $^1$S$_0$-$^3$P$_1$
transition is 163\,817.4\,(0.2)\,kHz. This value represents an improvement in
accuracy of more than 3 orders of magnitude with respect to previously
available data \cite{IsotopicShift}. The $^{86}$Sr optical frequency then
amounts to 434\,828\,957\,494\,(10)\,kHz.

The isotope-shift experiment provides also an indication of the
stability of the lock to the atomic line for periods longer than
2\,s \cite{UltraStableCavity}. We conclude that the observed
flicker noise at $5\times 10^{-12}$ in the absolute frequency
measurement may be attributed to the optical frequency comb
including its frequency reference. Moreover the relative
uncertainty of $1.2\times 10^{-11}$ due to uncontrolled systematic
effects does not explain completely the data scatter of $5\times
10^{-11}$ in the absolute frequency measurement. We did not
evaluate the noise performance in the GPS disciplined quartz
oscillator that is our local frequency reference. Possible sources
of noise are oscillation frequency sensitivity of quartz to
vibration and the behaviour of the complex, adaptive filter used
to discipline the quartz local oscillator to the GPS signal in the
10$^3$-10$^4$\,s region, which is the time period in which we are
making our measurements.

In conclusion, we demonstrated locking of a laser-diode to the visible
5s$^2$\,$^1$S$_0$-5s\,5p$^3$P$_1$ intercombination line of Sr and measured its
frequency using an optical-frequency comb-generator referenced to the SI second
through a GPS signal. The optical frequency measurement is obtained with a
relative uncertainty of $2.3\times 10^{-11}$, which represents an improvement
of more than 4 orders of magnitude with respect to previous data
\cite{Tino1994}. We also obtain an accurate value for the $^{88}$Sr-$^{86}$Sr
isotope shift improving the accuracy by more than 3 orders of magnitude.

Future improvements and developments involve cooling and trapping
of Sr atoms. Using cold atoms a precision in the range of one part
in 10$^{14}$ in one second can be expected with the transition
investigated in this work. Probing the ultra-narrow $0-0$ or $0-2$
transitions in cold trapped atoms should lead to a dramatic
improvement in stability and accuracy opening the way to the
$10^{-17}-10^{-18}$ range. A Sr-based optical reference could
employ all-solid-state laser sources (including light at 461\,nm
required for cooling and trapping). The realization of
ultra-precise optical frequency standards based on compact and
eventually transportable systems will enable future tests of
fundamental physics on Earth and in space.

We are grateful for the experimental assistance of T. Brzozowski
and C. De Mauro. We thank P. Lemonde for a critical reading of the
manuscript. G.M.T. also acknowledges seminal discussions with J.L.
Hall and C. Salomon. This work was supported by MIUR, EC (contract
No. HPMT-CT2000-00123), ASI and INFM.

\begin{table}
 \caption{
 \label{TabBiasUncertainty}
Budget of corrections and uncertainties for the $^{88}$Sr optical frequency
measurement; all values are in kHz.}

 \begin{tabular}{lrl}
   \hspace{1cm} {\sc Statistical value}  &         434\,829\,121\,316.5&(5.0)   \\ [0.5ex]\hline
                                &                \\   [-1.5ex]
1$^{\rm st}$ order Doppler     &0&(2)  \\
Recoil and 2$^{\rm nd}$ order Doppler      &-5.6&(0.1)      \\
2$^{\rm nd}$ order Zeeman      &-0.006&(0.003)   \\
Collisional shift              &0&(0.05) \\
Spectral purity                 &0&(0.5)  \\
Integrator offset               &0&(0.2) \\
Curvature and unbalanced intensity & 0&(2) \\[0.5ex]\hline
                                &         &       \\   [-1.5ex]
 \hspace{1cm}  {\sc Final value}             &       434\,829\,121\,311&(10)         \\
 \end{tabular}
\end{table}


\begin{thebibliography}{99}

\bibitem{Udem99}
Th.\,Udem, J.\,Reichert, R.\,Holzwarth, and T.W.\,H\"{a}nsch, Opt. Lett. {\bf
24}, 881 (1999).

\bibitem{Diddams2000}
S.A.\,Diddams {\it et al.}, Phys. Rev. Lett. {\bf 84}, 5102 (2000).

\bibitem{justifocation optical clocks}
Th.\,Udem, R.\,Holzwarth, and T.W.\,H\"{a}nsch, Nature {\bf 416}, 233 (2002).

\bibitem{HgCaClock}
T.\, Udem {\it et al.}, Phys. Rev. Lett {\bf 86}, 4996 (2001) and references
therein.

\bibitem{Hall89}
J.L.\,Hall, M.\,Zhu, P.\,Buch, J. Opt. Soc. Am. B {\bf 6}, 2194
(1989).

\bibitem{Katori00}
H.\, Katori {\it et al.}, in {\it Atomic Physics 17}, E.\,Arimondo,
P.\,De\,Natale, M.\,Inguscio Eds. (AIP, New York, 2001).

\bibitem{Tino1994}
G.M.\,Tino {\it et al.}, Appl. Phys. B {\bf 55}, 397 (1994).

\bibitem{Katori99}
H.\,Katori, T.\,Ido, Y.\,Isoya, and M.\,Kuwata-Gonokami, Phys. Rev. Lett. {\bf
82}, 1116 (1999).

\bibitem{Xu03}
X.\,Xu {\it et al.}, Phys. Rev. Lett. {\bf 90}, 193002 (2003).

\bibitem{SrTf03}
T.\,Mukaiyama {\it et al.}, Phys. Rev. Lett. {\bf 90}, 113002 (2003).

\bibitem{Courtillot03}
I.\,Courtillot {\it et al.}, to appear in Phys. Rev. A, preprint
arXiv:physics/0303023.

\bibitem{Katori03}
M.\,Takamoto and H.\, Katori, submitted, arXiv:physics/0309044.

\bibitem{Derevianko01}
A.\,Derevianko, Phys. Rev. Lett. {\bf 87}, 23002 (2001).

\bibitem{Dinneen98}
T.P.\,Dinneen {\it et al.}, Phys. Rev. A {\bf 59}, 1216 (1998).

\bibitem{Julienne}
A.\,Derevianko {\it et al.}, Phys. Rev. Lett. {\bf 90}, 063002 (2003).

\bibitem{Pound-Drever-Hall}
R.W.P.\,Drever {\it et al.}, App. Phys. B {\bf 31}, 97 (1983).

\bibitem{ClaironStabilization}
L.\,Hilico, D.\,Touahri, F.\,Nez, and A.\,Clairon, Rev. Sci.
Instr. {\bf 65}, 3628 (1994).

\bibitem{UltraStableCavity}
B.C.\,Young, F.C.\,Cruz, W.M.\,Itano, and J.C.\,Bergquist, Phys. Rev. Lett.
{\bf 82}, 3799 (1999).

\bibitem{Capillary}
I.\,Courtillot {\it et al.}, Opt. Lett. {\bf 28}, 468 (2003).

\bibitem{LaserLinewidth}
Sweeping the laser at 100\,Hz across the $^{88}$Sr resonance with
the reference cavity unlocked, or with the cavity locked to
$^{88}$Sr and sweeping across the $^{86}$Sr resonance at lower
frequency.

\bibitem{Cancio03}
F.\,Minardi {\it et al.}, Phys. Rev. A {\bf 60}, 4164 (1999).
M.\,Artoni, I.\,Carusotto, and F.\,Minardi, Phys. Rev. A {\bf 62},
023402 (2000).

\bibitem{Borde76}
J.L.\,Hall and C.J.\,Bord\'e, App. Phys. Lett. {\bf 29}, 788
(1976).

\bibitem{CollBroadening}
J.K.\,Crane, M.J.\,Shaw, and R.W.\,Presta, Phys. Rev. A {\bf 49}, 1666 (1994).

\bibitem{IsotopicShift}
F.\,Buchinger {\it et al.}, Phys. Rev. C. {\bf 32}, 2058 (1985).

\end{thebibliography}
\end{document}